# Unusual photoinduced crystal structure dynamics in TaTe$_2$ with double zigzag chain superstructure

J. Koga[1], Y. Chiashi[1], A. Nakamura[2], T. Akiba[1], H. Takahashi[3],

T. Shimojima[2], S. Ishiwata[3], K. Ishizaka[1,2]

[1]*Quantum-Phase Electronics Center and Department of Applied Physics, The University of Tokyo, Bunkyo, Tokyo 113-8656, Japan*

[2] *RIKEN Center for Emergent Matter Science (CEMS), Wako, Saitama, 351-0198, Japan*

[3]*Graduate School of Engineering Science, Osaka University, Toyonaka, Osaka 560-0043, Japan*

Email: ishizaka@ap.t.u-tokyo.ac.jp

**Abstract**

Transition metal dichalcogenides with superperiodic lattice distortions have been widely investigated as the platform of ultrafast structural phase manipulations. Here we performed ultrafast electron diffraction on room-temperature TaTe$_2$, which exhibits peculiar double zigzag chain pattern of Ta atoms. From the time-dependent electron diffraction pattern, we revealed a photoinduced change in the crystal structure occurring within <0.5 ps, though there is no corresponding high-temperature equilibrium phase. We further clarified the slower response (~1.5 ps) reflecting the lattice thermalization. Our result suggests the unusual ultrafast crystal structure dynamics specific to the non-equilibrium transient process in TaTe$_2$.

Transition metal dichalcogenide (TMD) $MX_2$ ($M$: transition metal, $X$: chalcogen) has been extensively studied from the viewpoint of fascinating photoinduced responses in ultrafast timescale. TMD consists of layered structures with weak stacking interaction. One well-known polytype is so-called $1T$ structure of $P\bar{3}m1$ space group [1], in which each layer consisting of the edge-sharing $MX_6$ octahedra forms a regular trigonal lattice. Depending on $M$ and $X$, $1T$-$MX_2$ shows various physical properties [2]. In particular, the formation of charge density wave (CDW) and corresponding superlattice distortion results in the divergence of electron-lattice coupled phenomena, including Mott transition, superconductivity, *etc*. [3,4]. Such CDW and superstructures in $MX_2$ are flexible to external triggers in many cases and regarded as one of the promising keys for ultrafast optical phase controls. For example, photoinduced phase transition in Mott insulator $TaS_2$ [5], topological switching in Weyl semimetal $WTe_2$ [6], and anomalous shear wave generation in $VTe_2$ [7,8] have been reported recently.

Among TMD, $TaX_2$ are representative systems that show characteristic superstructures and related photoinduced dynamics. It is known that $TaS_2$ shows multiple CDW states accompanying the so-called Stars of David type lattice distortions, as shown in Fig. 1(a) [9,10]. Recent development of ultrafast electron diffraction/microscopy enabled to visualize cooperative atomic motions occurring in the optical suppression process of CDW [5,11]. Also in $TaSe_2$, similar CDW [12] and photoinduced structural dynamics [13] were observed. In contrast, $TaTe_2$ shows a completely different lattice distortion as compared to $TaS_2$ and $TaSe_2$. At room temperature, $TaTe_2$ has a monoclinic crystal structure (space group C2/m, so-called $1T'''$-phase) as shown in Fig. 1(b), which is strongly distorted from $1T$ trigonal phase by forming the double zigzag chain of Ta running along the $b$-axis [14]. Such peculiar superstructure appears only in $MTe_2$ ($M$ = V,

Nb, Ta), and has been discussed based on the Fermi surface nesting scenario [15] and/or molecular-like trimerized bonding of *M* *d*-orbitals [16,17]. Whereas VTe$_2$ shows a structural phase transition from 1*T''* to the trigonal 1*T*-phase at 480 K [18,19], the undistorted 1*T*-phase in TaTe$_2$ has never been reported to the highest temperature. Nonetheless, TaTe$_2$ shows a structural phase transition to the low-temperature (LT) phase at 170 K where further clustering of Ta atoms into heptamer-pattern emerges [20]. In addition, a recent x-ray diffraction study has reported the anomalously large anisotropic displacement of Ta at the double zigzag chain center already at the room-temperature [21]. These indicate a characteristic structural instability inherent in 1*T'''*-TaTe$_2$, and unique photoinduced crystal structure dynamics can be expected. Although photoinduced dynamics in the LT phase has been reported recently [22,23], the room temperature 1*T'''*-TaTe$_2$ has not been investigated so far.

In this study, we perform ultrafast electron diffraction (UED) [24,25] on room temperature TaTe$_2$ to investigate the photoinduced structural dynamics in the 1*T'''*-phase. We successfully observed the ultrafast increase of some diffraction peak intensities, which indicates the change of crystal structure occurring within < 0.5 ps, though there are no corresponding higher temperature phases in equilibrium. We further investigate the transient structural dynamics in 1*T'''*-TaTe$_2$ by focusing on the time dependence profile.

Our UED system consists of a femtosecond pulsed laser (Light Conversion PHAROS) and an ultra-high-vacuum electron diffraction system with a photocathode (APCO ARH-30), as described in Ref. [26]. The diameters of the pump laser and probe electron beams are set to 300 and 120 µm at the sample, respectively, which ensure the homogeneous photoexcitation in the probing area. The time resolution of the measurements is set to 1 ps. All measurements for 1*T'''*-TaTe$_2$ are performed at room temperature (300 K), with a repetition rate of 10 kHz. The sample is photoexcited by a 1030 nm infrared optical pulse with 190 fs pulse duration. The pump fluence is set to 2.4 mJ/cm$^2$ unless otherwise noted. The maximum lattice temperature jump is estimated to be ~130 K from the specific heat capacity assuming the Dulong-Petit law and the optical absorption rate (44 %) of 60 nm TaTe$_2$ flake in Ref. [22]). The acceleration voltage of the electron diffraction system is set to 60 kV. Obtained 3-dimensional data (time-dependent diffraction pattern) is analyzed and visualized by lys software [27].

High-quality 1$T'''$-TaTe$_2$ single-crystals were grown by the chemical vapor transport method [28]. Thin flakes of 1$T'''$-TaTe$_2$ were prepared by ultramicrotome, where the target value of the thickness was 60 nm. The thin flakes were placed on a copper grid for UED measurements. In this study, the incident electron beam was fixed perpendicular to the TaTe$_2$ layer, i.e., perpendicular to both *a* and *b* (Fig. 1(c)). We use same orientation for the lattice vectors, *a*, *b*, and *c* as the literature [29], to describe the Miller indices of the diffraction spots.

Diffraction simulation of 1$T'''$-TaTe$_2$ was performed based on kinematic diffraction theory, by using the crystal structure in literature [20]. To consider the inhomogeneous bending of the flake sample with in the probing area, we introduce Gaussian-type broadening of the diffraction angle with a standard deviation of 3° following Ref. [7], by which the simulation pattern becomes similar to the experiment pattern.

Figure 2(a) shows the diffraction pattern of 1$T'''$-TaTe$_2$ before photoexcitation. The 2-fold symmetric diffraction pattern reflects the monoclinic space group (C2/m) of 1$T'''$-TaTe$_2$. We compared this pattern with the simulation in Fig. 2(b) (see method for detail), and confirmed the sufficient agreement. Figures 2(c-f) show the images of the diffraction patterns at time $t$ = -1.2 ps, 0.3 ps, 0.6 ps, 3.6 ps, subtracted by the one recorded at $t < 0$. In the lower-left inset of each panel, the magnification of the area depicted by the dashed rectangle is also shown. At $t$ = 0.3 and 0.6 ps [Fig. 2(d,e)], several diffraction peaks increase in intensity, which can be clearly seen in the inset. However, at 3.6 ps [Fig. 2(f)], diffraction intensities of all Miller indices decrease as compared to $t < 0$. At the same time, we observe the increase in the background intensity at 3.6 ps. Thus, we can find at least two different $t$-dependent behaviors in those diffraction patterns.

In the kinematic diffraction theory, a change in the diffraction intensity can be described either by the change of (1) lattice constants through the shape factor, (2) atomic coordinates through the structure factor, or (3) lattice temperature (Debye-Waller factor). Heating of the lattice temperature always decreases the Bragg diffraction intensities and increases the background intensity as the counterpart. Hence, the increase (decrease) of the background (diffraction) intensity observed at 3.6 ps can be interpreted as a consequence of the optically induced lattice thermalization. On the other hand, the increase of the diffraction peaks at $t$ = 0.3 ~ 0.6 ps cannot be explained by the photo-thermalization of lattice, thus raising the possibilities of the change in the shape factor and/or structure factor. In the present case of the 60 nm-thick flake sample, the characteristic time scale of the whole lattice constant modification (*i.e.* global lattice deformation) can be estimated by the sound velocity and the thickness, which should be in the order of tens of picoseconds [8,30–32]. This is significantly slow as compared to the observed $t$-dependent diffraction intensity increase. Thus, the present result indicates that the ultrafast photoinduced change in <1 ps should be reflecting the structure factor *i.e.* atomic coordinates.

We further discuss the origin of the photoinduced change in the crystal structure based on the temporal profile of diffraction intensities. Figure 2(g) shows the time dependences of 9 1 $\bar{3}$ and 11 1 $\bar{4}$ diffraction [orange arrows in Fig. 2(a)] and the background [solid white rectangle in Fig. 2(a)] intensities. Here we take the normalized intensity in the form of $I(t)/I(t<0)$, where $I(t)$ is the obtained intensity at *t*. The background intensity gradually evolves in $0<t<2$ ps, indicating the process of the lattice photo-thermalization. Its time profile can be analyzed by a curve fitting as follows. We assume that $I(t)/I(t<0)$ for the background can be described by the single exponential function $1+f(t;A,\tau)$ with:

$$f(t;A,\tau) = A\left(1-\exp\left(-\frac{t}{\tau}\right)\right)\theta(t),$$

where *A* and $\tau$ are the fitting parameters corresponding to the amplitude of photoinduced change and its time constant, respectively, and $\theta(t)$ is the Heaviside step function. To consider the 1 ps time resolution, we used $f(t)$ convolved with the Gaussian of 1 ps for the fitting. We determined the time constant for the lattice thermalization $\tau_{th} \sim 1.5$ ps by fitting the background intensity in Fig. 2(g). Regarding the 9 1 $\bar{3}$ diffraction intensity, we observe a slight increase at $t \sim 0$ to $\sim 0.6$ ps, consistent with Figs.2(d,e). This suggests that the observed structural change occurs within the time resolution. To confirm this scenario, we performed a further fitting analysis as follows. We assume that the 9 1 $\bar{3}$ diffraction intensity is described by two components: slow photo-thermalization of lattice and fast change in the crystal structure. Therefore, we use

$$1 + f(t;A_{th},\tau_{th}) + f(t;A_{st},\tau_{st}),$$

for the fitting, where $A_{th}$, $A_{st}$, and $\tau_{st}$ are the fitting parameters, whereas we fixed $\tau_{th} = 1.5$ ps. We again convolve this function with the 1 ps Gaussian function. The solid red curve in Fig. 2(g) shows that the 9 1 $\bar{3}$ diffraction intensity can be well fitted by this fitting function. We obtain upper limit of $\tau_{st} < 0.5$ ps ($\tau_{st} = 0.1$ ps is used in Fig. 2(g)). This upper limit indicates that the structural change occurs much faster than the lattice photo-thermalization. For the 11 1 $\bar{4}$ diffraction intensity, it only shows the decreasing behavior after the photoexcitation, unlike 9 1 $\bar{3}$. Nevertheless, it can be also reproduced by the similar fitting function with the same time constants ($\tau_{th}, \tau_{st}$), as shown in Fig. 2(g).

In addition, we investigate the fluence dependence of the photoinduced dynamics. Figures 3(b,c) show the fluence-dependent $I(t)/I(t<0)$ curves for the background and the 9 1 $\bar{3}$ diffraction intensity [the white rectangle and orange arrow in Fig. 3(a)]. The solid red curves in Fig. 3(b) are the fitting results with fixed $\tau_{th}$ (= 1.5 ps), which reproduce all the experimental data well. This means $\tau_{th}$ is independent of pump fluence. To estimate the fluence dependent $A_{th}$ and $A_{st}$, we performed the abovementioned fitting procedures in 9 1 $\bar{3}$ diffraction peak with fixed $\tau_{th}$ (= 1.5 ps) and $\tau_{st}$ (= 0.1 ps). Figures 3(d,e) show the resulting fluence dependence of $A_{th}$ and $A_{st}$ determine by the fitting. $A_{th}$, the decrease of the diffraction intensity with $\tau_{th}$ ~ 1.5 ps, monotonically evolves as the fluence is increased, reflecting the Debye Waller effect induced by photo-thermalization. Similarly, $A_{st}$ also shows a monotonic increase as a function of the fluence in this range. Generally in photoinduced phase transitions, it is commonly reported that threshold behavior and/or strong nonlinearity appear in the fluence dependence [33,8]. Also in VTe$_2$, the threshold behavior was observed at around 0.08 mJ/cm$^2$. However, considering that this is much smaller as compared to the lowest fluence of the present experiments (0.6 mJ/cm$^2$), it is difficult to rule out the existence of threshold. For further discussion, experiments in the lower fluence regime should be performed, though the measurement becomes difficult due to the weaker photoinduced response.

Finally, we discuss the mechanism of the possible change in the crystal structure of $1T'''$-TaTe$_2$. In TaTe$_2$, UED [22] and optical spectroscopy measurements [23] at LT phase have revealed the photoinduced structural change from low-temperature LT to high-temperature $1T''$ structure. In contrast, we note that the present measurements are performed at room temperature i.e. $1T''$-phase. As mentioned, there are no higher-temperature equilibrium phase reported so far. We also confirmed by the temperature dependent electron diffraction that no higher-temperature structural phase (such as $1T$) appears in our TaTe$_2$ sample up to 600 K, which is higher than the estimated maximum heating of sample (430 K). Therefore, a possible hidden state, which does not exist in thermal equilibrium, might be appearing in the strongly excited nonequilibrium process. For the room temperature $1T'''$-TaTe$_2$, we can hypothetically consider the structural change from $1T''$ to $1T$-phase, which is realized in the case of VTe$_2$ [7,8,18,19,29,34]. Nevertheless, we should note that in VTe$_2$, much lower fluence of 0.2 mJ/cm$^2$ can induce the structural change as large as > 10 % in diffraction intensity, whereas $A_{st} < 2\%$ at 2.4 mJ/cm$^2$ for the present case. We can also consider the possibility of yet unknown hidden states other than $1T$-phase. In a recent high-pressure x-ray study for TaTe$_2$, appearance of a superconductivity phase following a change in monoclinic angle was reported at low temperature (< 4 K) [4]. In addition, as mentioned above, some structural instability inherent to TaTe$_2$ was also reported at room temperature very recently [21]. In general, effect of electron multiple scattering and unintended local strain in the sample prevents the quantitative analysis of crystal structure from electron diffraction experiments. Therefore, development of novel quantitative crystal structure analysis technique in an ultrafast nonequilibrium regime overcoming the above problems is crucial for further understanding of the unusual crystal structure dynamics in TaTe$_2$.

In conclusion, we performed UED measurements for 1$T''$-TaTe$_2$ to investigate photoinduced structural dynamics. From the observed increase in the diffraction intensity, we elucidated the ultrafast crystal structural change occurring within 0.5 ps. By closely analyzing the time profile of the diffraction intensity, we found that the timescale of the lattice photo-thermalization was 1.5 ps, and could be safely ruled out as the origin of the structural change. To discuss the possibility of the hidden phase, quantitative crystal structure determination of ultrafast nonequilibrium state should be required for further investigation.

**Acknowledgments**

This work was partly supported by the JSPS KAKENHI (Grants No. JP19H05826, and No. JP22H00107). J. Koga acknowledges support by QSTEP at the University of Tokyo. We thank T. Nemoto for the thin-flake preparation.

**References**

[1] S. Manzeli, D. Ovchinnikov, D. Pasquier, O. V. Yazyev, and A. Kis, Nat. Rev. Mater. **2**, 1 (2017).
[2] J. A. Wilson and A. D. Yoffe, Adv. Phys. **18**, 193 (1969).
[3] B. Sipos, A. F. Kusmartseva, A. Akrap, H. Berger, L. Forró, and E. Tutiš, Nat. Mater. **7**, 960 (2008).
[4] J. Guo, C. Huang, H. Luo, H. Yang, L. Wei, S. Cai, Y. Zhou, H. Zhao, X. Li, Y. Li, K. Yang, A. Li, P. Sun, J. Li, Q. Wu, R. J. Cava, and L. Sun, Phys. Rev. Mater. **6**, L051801 (2022).
[5] M. Eichberger, H. Schäfer, M. Krumova, M. Beyer, J. Demsar, H. Berger, G. Moriena, G. Sciaini, and R. J. D. Miller, Nature **468**, 799 (2010).
[6] E. J. Sie, C. M. Nyby, C. D. Pemmaraju, S. J. Park, X. Shen, J. Yang, M. C. Hoffmann, B. K. Ofori-Okai, R. Li, A. H. Reid, S. Weathersby, E. Mannebach, N. Finney, D. Rhodes, D. Chenet, A. Antony, L. Balicas, J. Hone, T. P. Devereaux, T. F. Heinz, X. Wang, and A. M. Lindenberg, Nature **565**, 61 (2019).
[7] A. Nakamura, T. Shimojima, M. Matsuura, Y. Chiashi, M. Kamitani, H. Sakai, S. Ishiwata, H. Li, A. Oshiyama, and K. Ishizaka, Appl. Phys. Express **11**, 092601 (2018).
[8] A. Nakamura, T. Shimojima, Y. Chiashi, M. Kamitani, H. Sakai, S. Ishiwata, H. Li, and K. Ishizaka, Nano Lett. **20**, 4932 (2020).
[9] J. A. Wilson, F. J. Di Salvo, and S. Mahajan, Adv. Phys. **24**, 117 (1975).
[10] P. Fazekas and E. Tosatti, Philos. Mag. B (1979).
[11] S. A. Reisbick, Y. Zhang, J. Chen, P. E. Engen, and D. J. Flannigan, J. Phys. Chem. Lett. **12**, 6439 (2021).
[12] J. A. Wilson, F. J. Di Salvo, and S. Mahajan, Phys. Rev. Lett. **32**, 882 (1974).
[13] S. Sun, L. Wei, Z. Li, G. Cao, Y. Liu, W. J. Lu, Y. P. Sun, H. Tian, H. Yang, and J. Li, Phys. Rev. B **92**, 224303 (2015).
[14] B. E. Brown, Acta Crystallogr. **20**, 264 (1966).
[15] C. Battaglia, H. Cercellier, F. Clerc, L. Despont, M. G. Garnier, C. Koitzsch, P. Aebi, H. Berger, L. Forró, and C. Ambrosch-Draxl, Phys. Rev. B **72**, 195114 (2005).
[16] M. H. Whangbo and E. Canadell, J. Am. Chem. Soc. **114**, 9587 (1992).
[17] N. Mitsuishi, Y. Sugita, M. S. Bahramy, M. Kamitani, T. Sonobe, M. Sakano, T. Shimojima, H. Takahashi, H. Sakai, K. Horiba, H. Kumigashira, K. Taguchi, K. Miyamoto, T. Okuda, S. Ishiwata, Y. Motome, and K. Ishizaka, Nat. Commun. **11**, 2466 (2020).
[18] K. D. Bronsema, G. W. Bus, and G. A. Wiegers, J. Solid State Chem. **53**, 415 (1984).
[19] T. Ohtani, K. Hayashi, M. Nakahira, and H. Nozaki, Solid State Commun. **40**, 629 (1981).
[20] T. Sörgel, J. Nuss, U. Wedig, R. Kremer, and M. Jansen, Mater. Res. Bull. **41**, 987 (2006).
[21] N. Katayama, Y. Matsuda, K. Kojima, T. Hara, S. Kitou, N. Mitsuishi, H. Takahashi, S. Ishiwata, K. Ishizaka, and H. Sawa, Phys. Rev. B **107**, 245113 (2023).
[22] K. M. Siddiqui, D. B. Durham, F. Cropp, C. Ophus, S. Rajpurohit, Y. Zhu, J. D.

Carlström, C. Stavrakas, Z. Mao, A. Raja, P. Musumeci, L. Z. Tan, A. M. Minor, D. Filippetto, and R. A. Kaindl, Commun. Phys. **4**, 152 (2021).

[23] T. C. Hu, Q. Wu, Z. X. Wang, L. Y. Shi, Q. M. Liu, L. Yue, S. J. Zhang, R. S. Li, X. Y. Zhou, S. X. Xu, D. Wu, T. Dong, and N. L. Wang, Phys. Rev. B **105**, 075113 (2022).

[24] A. H. Zewail, Annu. Rev. Phys. Chem. **57**, 65 (2006).

[25] B. J. Siwick, J. R. Dwyer, R. E. Jordan, and R. J. D. Miller, Chem. Phys. **299**, 285 (2004).

[26] A. Nakamura, T. Shimojima, M. Nakano, Y. Iwasa, and K. Ishizaka, Struct. Dyn. **3**, 064501 (2016).

[27] A. Nakamura, J. Open Source Softw. **8**, 5869 (2023).

[28] N. Mitsuishi, Y. Sugita, T. Akiba, Y. Takahashi, M. Sakano, K. Horiba, H. Kumigashira, H. Takahashi, S. Ishiwata, Y. Motome, and K. Ishizaka, Phys. Rev. Res. **6**, 013155 (2024).

[29] T. Suzuki, Y. Kubota, N. Mitsuishi, S. Akatsuka, J. Koga, M. Sakano, S. Masubuchi, Y. Tanaka, T. Togashi, H. Ohsumi, K. Tamasaku, M. Yabashi, H. Takahashi, S. Ishiwata, T. Machida, I. Matsuda, K. Ishizaka, and K. Okazaki, Phys. Rev. B **108**, 184305 (2023).

[30] M. Harb, W. Peng, G. Sciaini, C. T. Hebeisen, R. Ernstorfer, M. A. Eriksson, M. G. Lagally, S. G. Kruglik, and R. J. D. Miller, Phys. Rev. B **79**, 094301 (2009).

[31] L. Wei, S. Sun, C. Guo, Z. Li, K. Sun, Y. Liu, W. Lu, Y. Sun, H. Tian, H. Yang, and J. Li, Struct. Dyn. **4**, 044012 (2017).

[32] A. Nakamura, T. Shimojima, and K. Ishizaka, Nano Lett. **23**, 2490 (2023).

[33] E. Möhr-Vorobeva, S. L. Johnson, P. Beaud, U. Staub, R. De Souza, C. Milne, G. Ingold, J. Demsar, H. Schaefer, and A. Titov, Phys. Rev. Lett. **107**, 036403 (2011).

[34] H. Tanimura, N. L. Okamoto, T. Homma, Y. Sato, A. Ishii, H. Takamura, and T. Ichitsubo, Phys. Rev. B **105**, 245402 (2022).

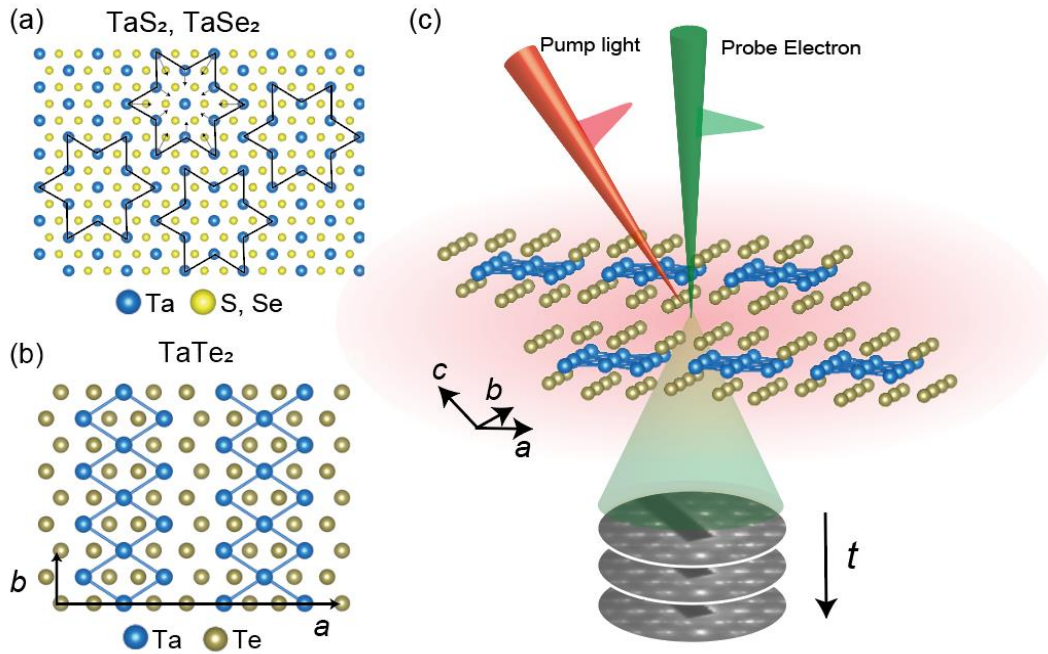

**Fig. 1.** In-plane crystal structures of TaS$_2$, TaSe$_2$ (a) and TaTe$_2$ (b). The black arrows in (b) denote the *a* and *b* axes. (c) Schematic of the experimental setup. We homogeneously excite the 1*T''*-TaTe$_2$ flake sample by 1030 nm pulsed light, and the electron diffraction measurements are performed by using the pulsed electron beam. The electron beam is set perpendicular to the *a* and *b* axes of 1*T''*-TaTe$_2$.

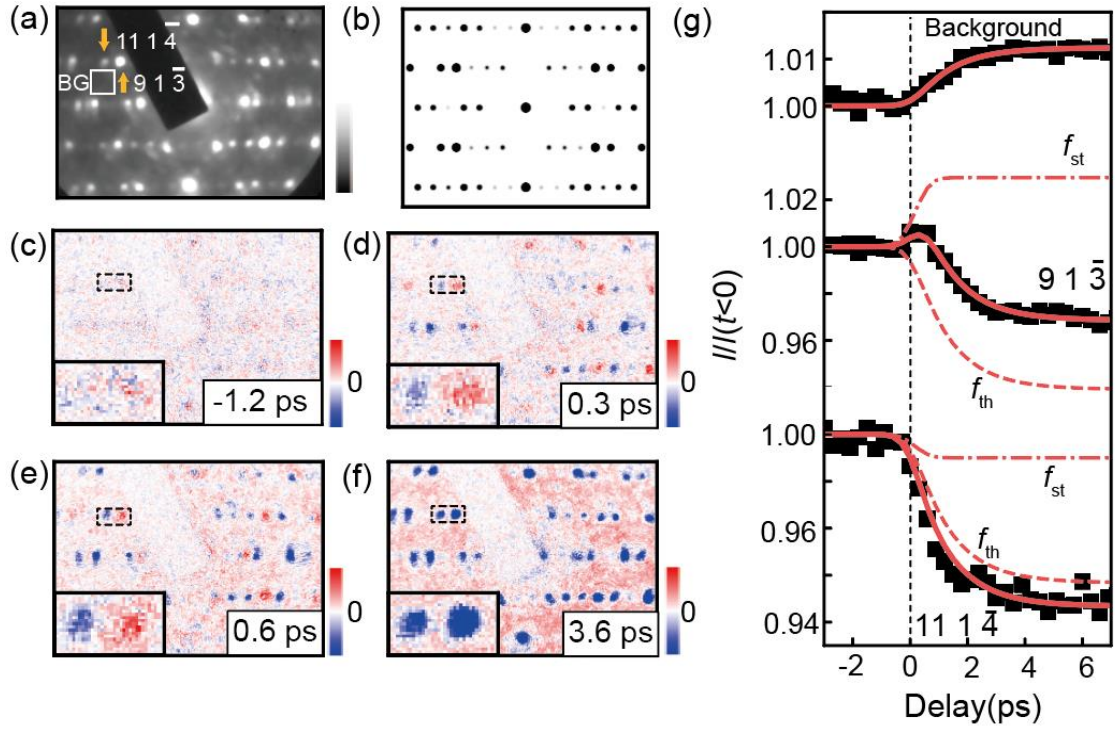

**Fig. 2.** (a) Electron diffraction pattern of $1T''$-TaTe$_2$ obtained before photoexcitation. (b) Simulated electron diffraction pattern of $1T''$-TaTe$_2$. (c-f) Time-dependent change in diffraction pattern recorded at -1.2, 0.3, 0.6, and 3.6 ps, respectively. The diffraction pattern before photoexcitation is subtracted to show the increase and decrease in the red-white-blue color scale. (g) Time-dependent normalized intensity of background and 9 1 $\bar{3}$, 11 1 $\bar{4}$ Bragg diffractions. For the background intensity, we integrate the intensity in the solid white rectangle in (a), which is denoted as BG. The solid red curves denote the fitting results. $f_{th}$ and $f_{st}$ denote the curves corresponding to $f(t; A_{th}, \tau_{th})$ and $f(t; A_{st}, \tau_{st})$, respectively.

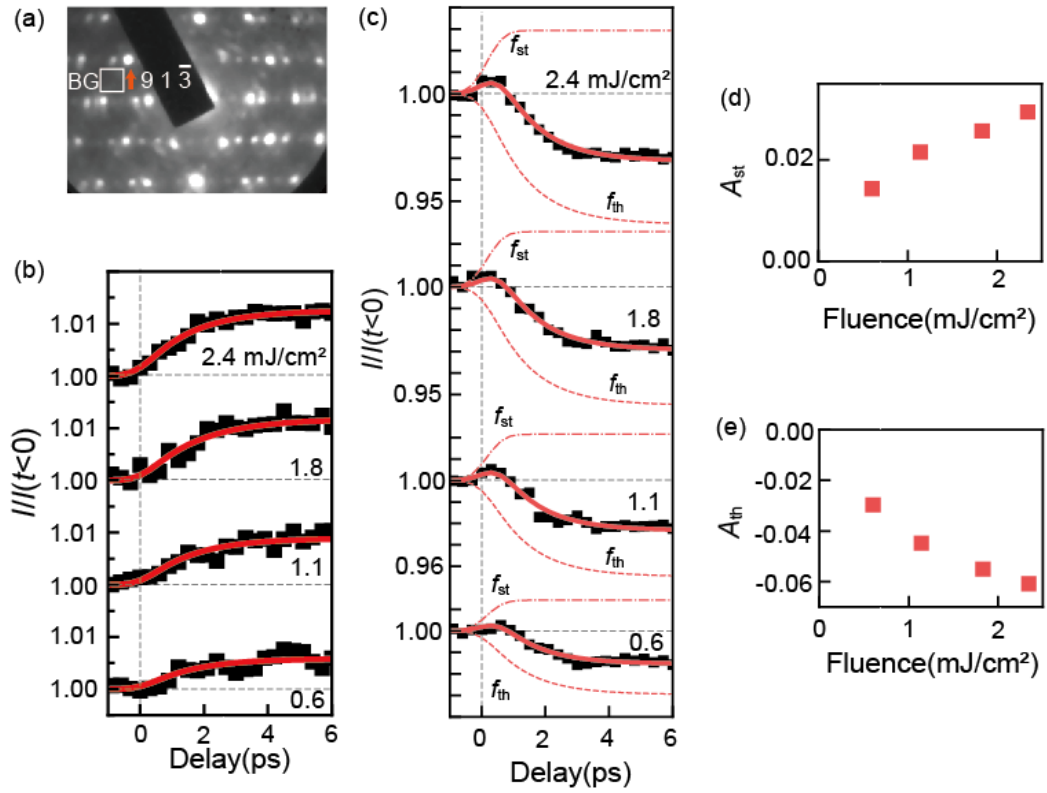

**Fig. 3.** (a) Electron diffraction pattern of $1T''$-TaTe$_2$. The white rectangle and orange arrow indicates the background intensity and the $9\,1\,\bar{3}$ Bragg diffraction, respectively. (b,c) Normalized diffraction intensity of background intensity and $9\,1\,\bar{3}$ Bragg peak obtained for 0.6, 1.1, 1.8 and 2.4 mJ/cm$^2$ pump fluences. The solid red curves denote the fitting result for respective data. $f_{\text{th}}$ and $f_{\text{st}}$ denote $f(t; A_{\text{th}}, \tau_{\text{th}})$ and $f(t; A_{\text{st}}, \tau_{\text{st}})$, respectively. (d-f) Fluence dependences of $A_{\text{st}}$ and $A_{\text{th}}$ obtained by fitting the experimental result in (c).